\documentclass[journal = ancac3, manuscript = article, layout = twocolumn]{achemso}
\usepackage{amsmath}
\usepackage[usenames,dvipsnames]{xcolor}

\title{Spin locking at the apex of nano-scale platinum tips}
\author{Richard Koryt\'ar}
\email{richard.korytar@kit.edu}
\affiliation[Institut f\"ur Nanotechnologie, Karlsruhe Institute of
Technology]{Institut f\"ur Nanotechnologie, Karlsruhe Institute of
Technology (KIT), D-76344 Eggenstein-Leopoldshafen, Germany}

\author{Ferdinand Evers}
\affiliation[Institut f\"ur Nanotechnologie, Karlsruhe Institute of
Technology]{Institut f\"ur Nanotechnologie, Karlsruhe Institute of
Technology (KIT),  D-76344 Eggenstein-Leopoldshafen, Germany}
\alsoaffiliation[Institut f\"ur Theorie der Kondensierten Materie, Karlsruhe
Institute of Technology]{ Institut f\"ur Theorie der Kondensierten Materie, Karlsruhe
Institute of Technology (KIT), D-76128 Karlsruhe, Germany}
\alsoaffiliation[DFG-Center for Functional Nanostructures, Karlsruhe
Institute of Technology]{DFG-Center for Functional Nanostructures, Karlsruhe
Institute of Technology (KIT), D-76131 Karlsruhe, Germany}

\begin{document}

\begin{abstract}
Nanostructures based on platinum, such as small clusters or STM-tips, 
often exhibit an atomistic structure that
relies upon one or very few strongly under-coordinated 
platinum atoms. Here, we analyze a paradigmatic example, 
an apex atom on a pyramidal platinum cluster employing the 
density functional theory. We show that such a pristine platinum tip 
exhibits a spin polarization of the apex atom with a remarkable
robustness. Due to a depletion of the 
projected density of states at the apex position, 
the apex-magnetization is efficiently locked to about 0.6$\mu_\text{B}$.
\end{abstract}


\begin{tocentry}
\begin{minipage}{0.5\textwidth}
\includegraphics[height=4.0cm]{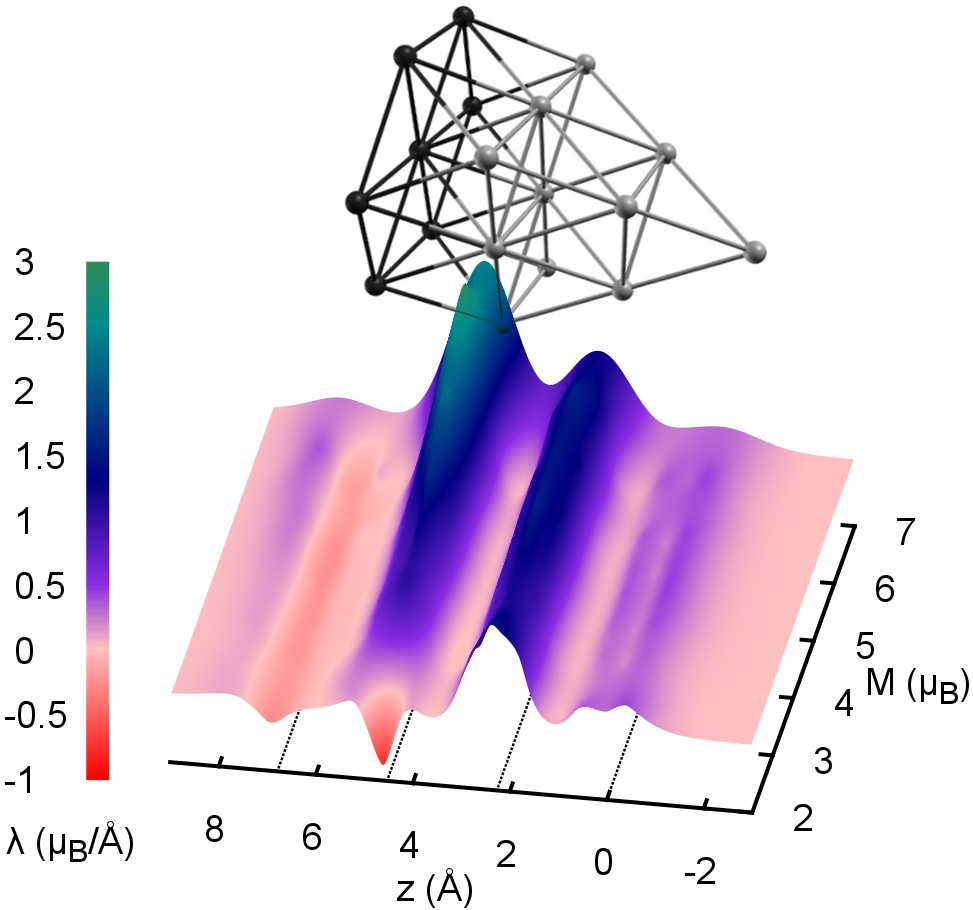}
\end{minipage}
\begin{minipage}{0.5\textwidth}
The spin density on the apex atom of a pyramid-shaped platinum tip is 
insensitive to changes of the total magnetization and stays locked
to about $0.6\mu_B$.
\end{minipage}
\end{tocentry}

\section{INTRODUCTION}

Recent years have seen a significant amount of cross-fertilization  
between the research fields of spintronics and molecular electronics. 
And indeed, exploiting spin and charge at the same time could open a 
fascinating venue towards implementing \emph{memory} (spin) and 
\emph{logic} (charge) on the same device 
\cite{sanvito2011molecular,shiraishi2011molecular}.
For tailoring and controlling spin and charge transport through contacted nano-scale
devices (molecules, atomic chains) it is necessary, if the device properties 
can be well-defined down to the atomic scale. 
Therefore, a considerable effort in {\em Molecular Spintronics} is devoted 
to optimizing conditions so device properties become (1) reproducible and 
(2) well controlled.\cite{lathasStuff}
The proper choice of the contact 
material is a central part of this endeavor, because on the atomic level 
contacts are invasive, in general, and must be considered to be a part of the device. 

Here we report a result that is of relevance for the use of platinum as 
a nano-electrode material: due to the reduction of the coordination number a single 
platinum atom adsorbed on the bulk material exhibits a very 
strong tendency towards magnetism. 
 
As is well known, platinum is an important material from the technological point 
of view. Besides being a commonly used electrode in catalytic chemistry,
it has been employed in scanning tunneling microscopy measurements \cite{platinumTip}. 
Also, platinum has been tested as contact material in {\em Molecular Electronics},
especially in break junction experiments \cite{PaulyPrl08,C2NR30832K}, 
since like gold it forms mono-wires when 
pulling the junction\cite{agrait:pr} even if bridged by small molecules \cite{yelin2013}.
However quite unlike gold, platinum has intriguing magnetic properties that may offer 
qualitatively new possibilities for contact design. They exist due to the proximity
to a Stoner instability. While in the bulk form
it is  known to be non-magnetic, platinum  develops super-paramagnetic
behavior in the case of long mono-wires\cite{Pietro} and small nanoclusters.

Platinum electrodes have been studied by several authors 
\cite{PalaciosPrb05,WangPrb11,SmogunovTosattiPrb08,
ThiessHeinzePrb10,SuarezFerrerPrb09}
in theoretical transport studies of nano-wires and 
break-junctions. Fern\'andez-Rossier \emph{et.~al}
\cite{PalaciosPrb05} 
showed that the finite magnetization of a short Pt chain
survives even after attaching it to pyramid-like electrodes. 
It is thus clear, that the reduction of the coordination 
number from three-dimensional (12) to one-dimensional (2) 
is sufficient for a magnetic transition to take place 
-- at least on the level of density-functional-theory, DFT, 
with conventional exchange-correlation functionals. 
The question, that we would like to address here is, whether 
a single apex-atom of such a pyramid by itself would be 
magnetic already. This would be an intermediate case with 
a coordination number of three or four, depending on 
the specifics of the contact pyramid.
We are motivated by the observation that such 
geometries are typical for applications of platinum as contact material 
in \emph{Molecular Electronics}.

Our study suggests that already the single apex-atom 
is magnetic (on DFT-level). Moreover, we see that the magnetization 
invokes a kind of proximity effect, so that it enters the pyramid 
down to first and second neighbor distances. To the best of our knowledge, 
such a tip magnetization has not been reported yet. 
In the reference \cite{SmogunovTosattiPrb08} the electrodes were
modeled by infinite surfaces, and no finite magnetization
seems to accumulate in the electrode volume. Thiess, \emph{et al.}
\cite{ThiessHeinzePrb10} used cluster-like electrodes in order to
simulate a break-junction condition, yet the penetration
of magnetization to the bulk was not reported by the authors.

Magnetism is treated in our work on DFT-level in the all-electron
full-potential approach. We study platinum pyramids enforcing 
full-surface conditions by applying non-magnetic (\emph{i.e.} non-polarized) 
boundary conditions on one of the faces and
varying the total magnetization in the remaining volume. 
We observe that the tip magnetization decays from the apex towards the 
non-polarized bulk, spanning several atomic sites. The apex magnetization
is insensitive to the details of the boundary conditions, indicating that
the magnetization of the tip stems from the geometric quantum confinement.

Quantum confinement has been discussed as a possible agent to drive
magnetic instabilities in itinerant electron systems before. 
We mention the 0.7 anomaly that can show up in the conductance of quantum point contacts 
\cite{PhysRevLett.77.135,PhysRevB.44.13549}
It was explained as a consequence of a reduction of screening in the 
contact region due to the confined geometry. The loss of screening enhances
electron interactions and thus can drive magnetic transitions, at least
in principle. 

Theoretical studies of the phenomenon 
approximate the electronic structure by a two dimensional
jellium \cite{rejec2006magnetic}, which is appropriate for quantum-point
contacts as realized in semiconductor technologies. By contrast, we
demonstrate the phenomenon in a realistic three-dimensional system (corner of 
a Pt crystal) by using atomistic first-principles calculations.
In the few-atom platinum clusters, the origin of spin magnetic moments can
be attributed to the potential of the core, giving rise to localized
$3d$ states with enhanced Coulomb interaction. Our claim is that the polarization
survives also in the bulk limit, in regions of geometric confinement
as tips, corners, necks or surface islands and corrugations.

We emphasize that the basic mechanism, quantum confinement and breakdown of screening, 
is not restricted to platinum and could apply to other metals close 
to magnetic transitions.
So far, an intrinsic electrode magnetization was not reported for 
iridium \cite{SuarezFerrerPrb09} and palladium junctions\cite{Smelova,TosattiPd},
mainly because electrodes were simulated by plain surfaces.
We suspect, that it could exist, nevertheless, 
for the physical reasons outlined in this manuscript.

\begin{figure}
\centering
\includegraphics[width=\columnwidth]{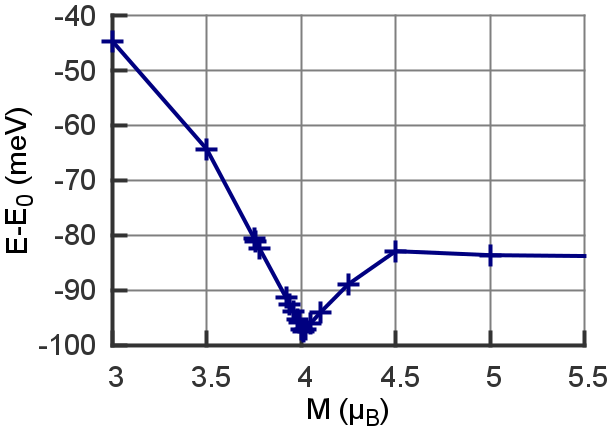}
\caption{\label{fig:totale}Total energy of the 18-atom pyramid 
as a function of the total magnetization.
Zero energy corresponds to the unpolarized solution.}
\end{figure}

\section{RESULTS}
\begin{figure}[htb]
\centering
\includegraphics[width=\columnwidth]{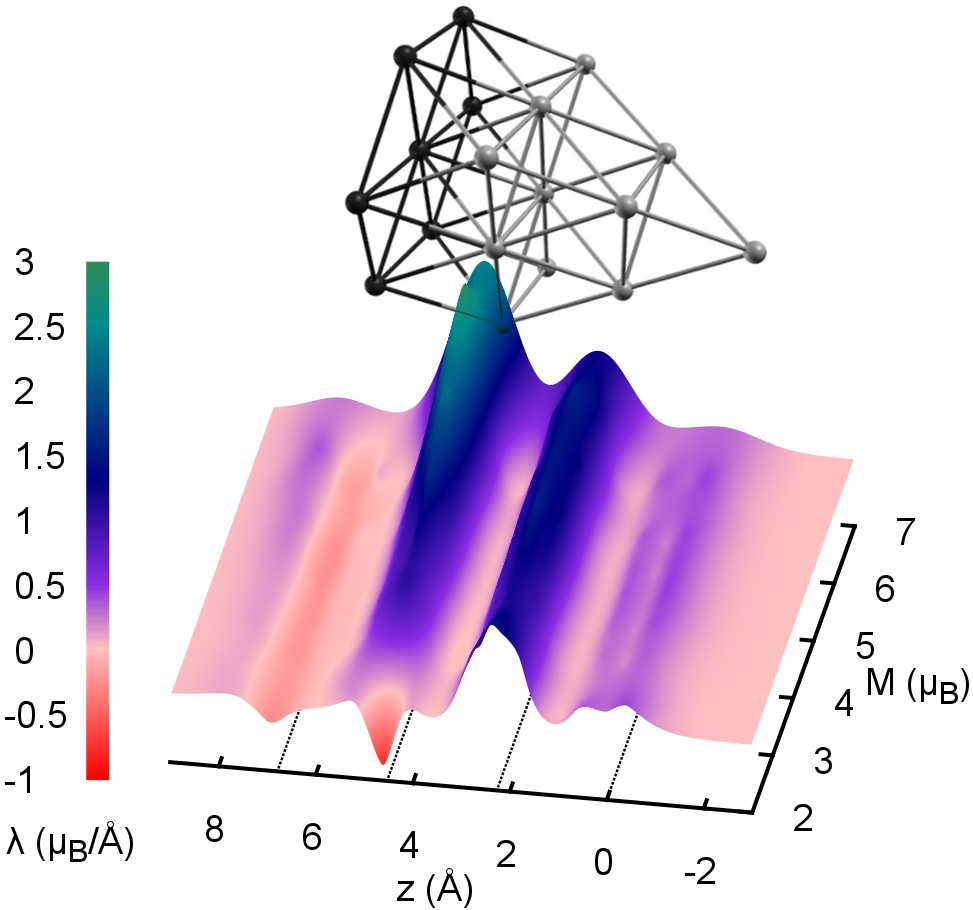}
\caption{\label{Fig:ixy18}Upper part:
Ball-and-stick model\cite{XCrysDen} of the pyramid. The platinum atoms whose magnetization
is constrained to vanish are darker.
Lower part:
Line spin density $\lambda$ along the $z$ axis of the 18-atom pyramid for
magnetization varying between M=2 and M=7 $\mu_B$.
Line spin density varies on the vertical axis and is color-coded for
clarity. Atomic layers are positioned at $z=0,2.27,4.53,6.79$\AA\
starting from apex and visualized by dashed lines in the horizontal plane.
}
\end{figure}

We have performed extensive density functional studies of platinum
pyramids of several sizes. Here we show results for a pyramid
of 18 platinum atoms, as in \ref{Fig:ixy18}. All physical
quantities are calculated in DFT subject to a boundary condition of
vanishing spin density on the pyramidal face opposite to the apex
(see Methods for the details of our calculations).

\ref{fig:totale} displays dependency of the total energy 
on the magnetization of the 18-atom Pt-cluster. 
We find a minimum energy for $M\approx 4.01\mu_B$. Naturally, this 
corresponds to a magnetization spanning several atomic sites.
To inspect the local behavior of spin density upon the variation of
magnetization, we calculate the line density
\begin{equation}
\label{e1}
\lambda (z) = \iint \left[n_{\uparrow}(\mathbf r)-n_{\downarrow}(\mathbf r)\right]
    \text{d}x\text{d}y
\end{equation}
where $n_{\uparrow,\downarrow}(\mathbf r)$ is the 
spin majority/minority electronic density. The $z$ axis goes through
the apex perpendicular to the interface layer.
\ref{Fig:ixy18} shows $\lambda (z)$ as $z$ goes through the pyramid.
Pink areas are regions of negligible spin-polarization. They are located
outside the pyramid, as well as in the interstitials, giving rise to
a ridge-valley picture. This reflects the fact that spin polarization is accumulated
in the compact $5d$ orbitals, rather than in the overlapping $sp$ orbitals.
Interestingly, the ridges at $z=0$ and 2.27\AA\ have
constant height independent on the total magnetization. The entire variation
of the magnetization is effectively restricted to the third layer (the third
ridge starts from red (negative $\lambda$) to end up in green.
Finally, the fourth layer is the constrained interface. Here only a small
artefact of spin polarization appears\bibnote{Due to applying the 
constraint on the atomic Mulliken polarization and not
on each orbital separately.}.

\begin{figure}
\centering
\includegraphics[width=\columnwidth]{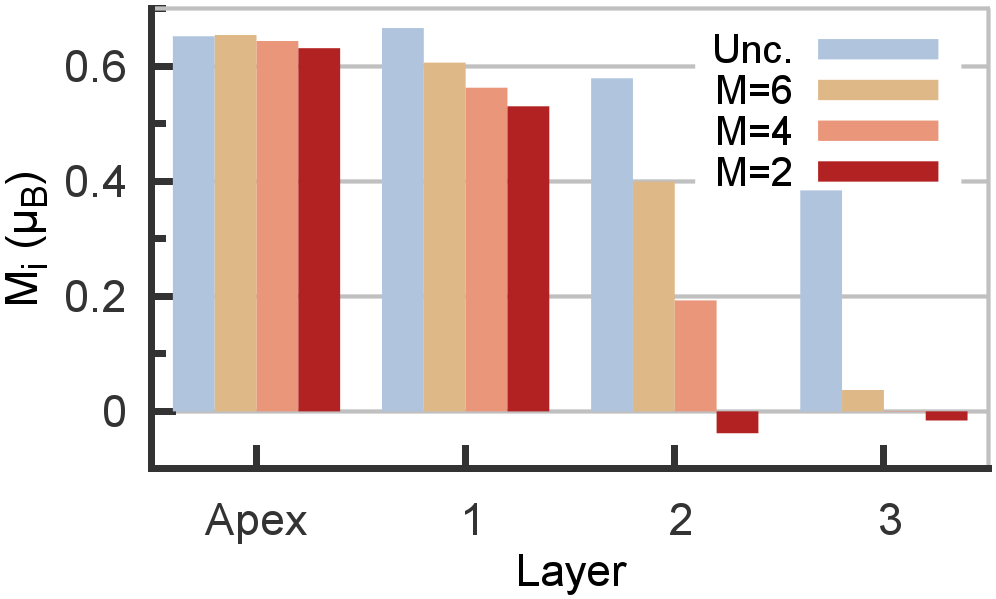}
\caption{\label{Fig:hist}Magnetic moment per atom in the layers 0-3
of the pyramid. Zero corresponds to the apex.}
\end{figure}

The ridge-valley structure allows unambiguous decomposition of magnetization to
atomic layers - simply by integrating the line density between interstitials.
The plot in \ref{Fig:hist} contains the layer magnetization divided by
the number of atoms in the layer.
Surprisingly, if the spin density is allowed to relax
freely (i.e. switching off the constraint at the interlayer), 
the highest polarization is located in the layer next to the apex.
If the constraint is applied, the magnetization
decays towards the interface region as expected.
\ref{Fig:hist} demonstrates again that the apex is almost insensitive
to the interface region (i.e. to the constraint). Change in the overall constraint
is compensated in the deeper layers of the pyramid. The robustness of the
apex indicates, that its magnetization is determined by quantum confinement,
i.e. geometry of the tip. 

\begin{figure}
\centering
\includegraphics[width=\columnwidth]{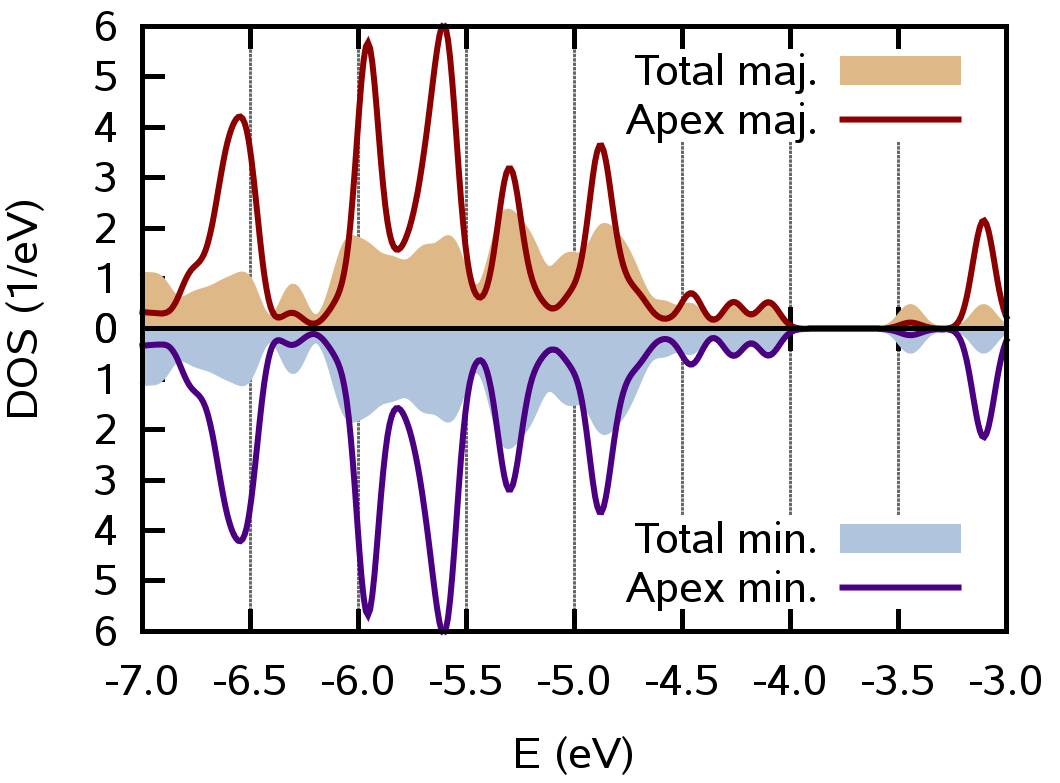}
\includegraphics[width=\columnwidth]{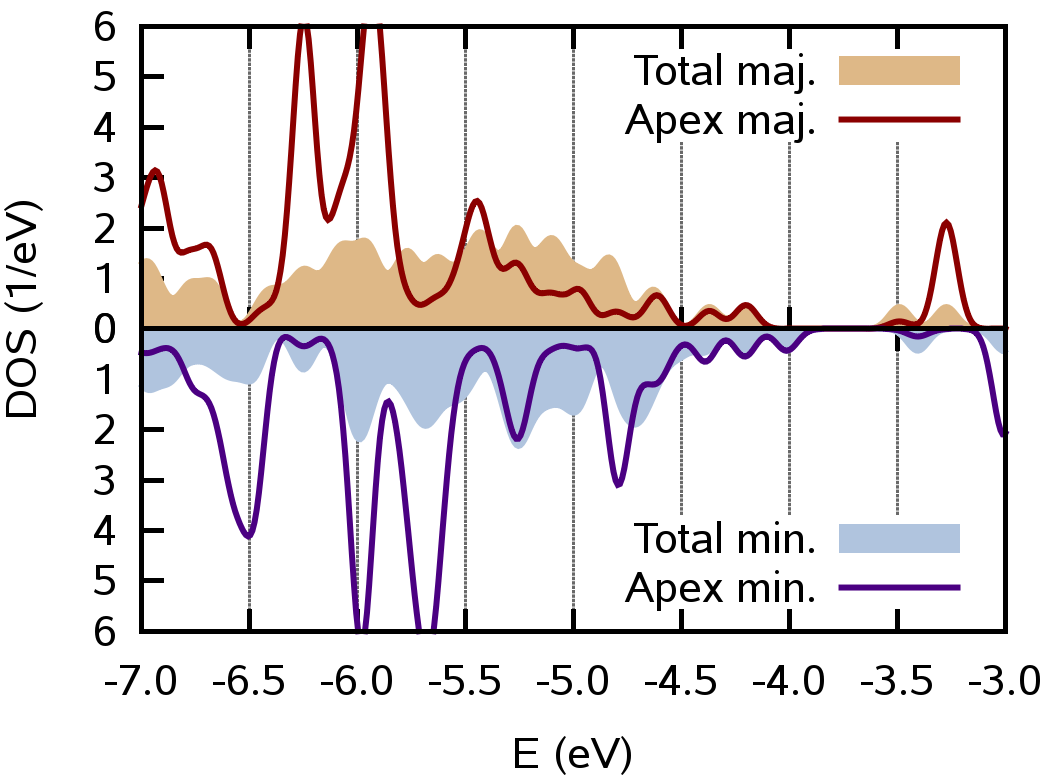}
\caption{\label{Fig:diff0}Total and apex-projected density of states of the 
18-atom pyramid. The total DOS has been divided by the number 
of atoms. The left plot
presents an unpolarized solution, the right plot presents polarized $M=4.01
\mu_B$. Upper (lower) panel shows majority (minority) sector. The approximate
chemical potential lies at $E=-4.89$ eV.}
\end{figure}

Finally, we  demonstrate that the tight locking of apex magnetization
may be understood as a consequence of depletion of apex density of states.
\ref{Fig:diff0} presents the density of states of the 
18-atom pyramid. 
\bibnote{The discrete spectrum has been convoluted with
Gaussian smearing $0.05$ eV, which roughly equals the average level spacing in the region of
interest. The chemical potential estimated as the average of highest
occupied and lowest unoccupied single-particle energy level, is $-4.89$ eV.}
In the unpolarized case, the region below the electro-chemical potential, $\mu$, shows higher spectral density,
contributed mainly by the $5d$ orbitals.
The chemical potential lies at the spectral resonance of a strong apex character.
When allowing for (constrained)
spin polarization, this spectral peak DOS is pushed to lower (higher) energies in the 
majority (minority) spin. At the same time, the apex density of states
around the chemical potential is depleted. Hence, small deviations
of the total magnetization can not be carried over in the apex region,
since available single-particle states have little weight there.

\section*{Discussion}
We discuss the possible impact of two approximations underlying our 
calculations. 

Firstly, we recall that we have to rely upon approximate functionals
for electronic exchange and correlation. 
Therefore, one expects that our results inherit typical artefacts of mean-field 
treatments. Such are related to underestimating quantum fluctuations so that 
broken-symmetry solutions are overemphasized. 
Still, we expect that the main features of our calculation survive 
also in the presence of quantum effects.

Our main argument is that the observed magnetic moment,
albeit localized, spans a few atomic sites,
and has a relatively large magnitude, 
($M\approx 4 \mu_b$), which is not easily Kondo-screened by the bulk electrons. 
Hence, one expects that a major fraction of the spin will persist down to
lowest temperatures.

However, there is a suspicion that this case is slightly different from the
commonly studied situation, 
where a transition-metal atom sits on top of a flat non-magnetic
substrate (no pyramid). Magnetic ad-atoms such as Co exhibit Kondo effect
\cite{kernCoPrl} 
due to anti-ferromagnetic hybridization exchange with the paramagnetic host.
In the case studied here, ferromagnetic-like instabilities at low-coordinated
platinum atoms require ferromagnetic interactions (like Hund's rule exchange).
Hence, the specific nature of quantum fluctuations calls for further 
investigations.
\bibnote{Tosatti \emph{et al.} 
\cite{gentile2009lack} have elaborated on the topic. They consider a nanoscale spin moment embedded
in a metal close to a ferromagnetic transition. For them the spin is necessarily subject to a ferromagnetic
exchange (interatomic direct Coulomb). The Coulomb-interaction competes with an anti-ferromagnetic
exchange due to hybridization (kinetic, Kondo exchange). To our knowledge, the scenario
when the ferromagnetic interaction pervades has not been observed yet.
 }
We remark that the situation here bears similarity to the local moment
formation observed in ferromagnetic atomic contacts 
\cite{Untiedt}. Iron, cobalt and nickel (1) form \emph{localized} magnetic moments in the
contact region, despite the \emph{itinerant} magnetism of these elements in bulk
(2) and these moments are subject to experimentally observed Kondoesque fluctuations.


Second, we have not been accounting for the spin-orbit interaction. 
Indeed, studies of platinum nano-structures 
suggest that a significant spin-orbit interaction could be present. 
For instance, Tosatti and coworkers \cite{smogunov2007colossal}
predicted collosal anisotropy of magnetization of infinite mono-wires.

To address the consequences of spin-orbit interactions for our results, we
recall that the main effect of this interaction will be to fix the direction of a 
given magnetic moment. Hence it helps to decrease quantum fluctuations and will 
favor apex magnetism working into the direction of our claim. 

\section*{CONCLUSIONS}
We have demonstrated that platinum tips develop sizable magnetization
($4\mu_B$)
even in the presence of a paramagnetic boundary. The spin density
at the apex is effectively locked due to quantum confinement, manifesting
itself in a local depletion of density of states. The magnetization per atom
decays from the apex to the neighboring layers, mimicking a magnetic proximity
effect. 

The very presence of magnetization must be kept in mind
whenever platinum electrodes are used in nano-scale transport
studies. More generally, our results apply to a broader class
of nano structures where, as we claim, low-coordinated platinum atoms
drive a local magnetic instability. The unexpected deviation 
from the \emph{normal metal} paradigm can have consequences important for 
molecular electronics and spintronics.

\section{METHODS}

We perform first-principles calculations based on spin-polarized
density-functional theory. 
\bibnote{The exchange correlation (XC) functional used is PBE \cite{PerdewPBE}.
The Kohn-Sham states are represented in an atom-centered  basis
set as implemented in the \textsc{Fhi-Aims} package \cite{AimsReference} fully
including all core electrons. Scalar relativity is included 
in the zero order regular approximation \cite{LentheZora}.
The basis functions are numerical and strictly localized.}
The focus is on platinum clusters of pyramidal shape
 ``cut'' from the face-centered crystal
of platinum, so that the axis points in the crystalline $(111)$ direction.
We use a family of pyramids of 11 atoms (3 layers), 18 atoms
(4 layers, see \ref{Fig:ixy18}) and 33 atoms (5 layers) to 
be able to extrapolate finite-size effects. The nearest-neighbor spacing is
2.7749 \AA. The threefold symmetry axis $(111)$ of the pyramid is 
lifted by including ad-atoms.
In order to enforce a non-magnetic boundary on the face opposite to the apex
(interface layer), the spin densities on the atoms of the interface
layer are forced to vanish. We achieve this by using a variant of the 
constrained density functional 
theory according to the reference \cite{SchefflerCdft}. 

We add further remarks on technicalities: 
In principle,
the constraint specification does not assure unique ground state
spin density and the choice of initial density matrix is crucial
for the convergence of magnetic systems. We have tested several initial 
spin configurations on the atoms and did not see evidence for
inter-atomic coupling of anti-ferromagnetic character.

To substantiate the impact of basis set size, we studied
a small 11-atom pyramid without constraints for parameter sets
\textsc{light/tier1, light/tier2, tight/tier1, tight/tier2} of the 
\textsc{Fhi-Aims} package\cite{AimsReference}. The total
magnetization does not differ more than a few percent.
Therefore we employ the basic basis set light/tier1 for most
of the numerical results. Larger basis sets involve
extended orbitals, these being a drawback in the constrained
calculations, involving diffusion of constrained regions.

Size effects have minor impact on the accumulated spin.
For the 11, 18, 33-atom pyramids we always find $M\approx 4\mu_B$ as
the energy minimum, although the distribution of spin
over layers changes.

We confirm the stability of the magnetic solution by checking 
against the GGA$+U$ functional~\cite{YuLDAU} (using $U=2$ eV).
In the 11-atom pyramid, we see a small decrease of the apex
magnetization from 0.69 to 0.60 $\mu_B$.

\begin{acknowledgement}
The authors thank A.~Bagrets for discussions and revision of
the manuscript. Funding by the DFG-Center of Functional Nanostructures
and by the DFG Priority Program 1243 is acknowledged. Numerics has
been done at HC3 cluster of the Karlsruhe Institute of Technology,
Steinbuch Center for Computing.
\end{acknowledgement}
\bibliographystyle{achemso}
\bibliography{references}
\end{document}